\documentclass[12pt]{book}

\usepackage[dvips]{graphicx,color}
\usepackage{makeidx,tsukuba}

\makeauthorindex
\makeindex

\newcommand\paperno{
   \vspace{-8\baselineskip} 
   \noindent \underline{\it LANL Report \rm \# LA-UR-03-3406}
   \vspace{6.8\baselineskip}}

\begin{document}

\BookTitle{\itshape The 28th International Cosmic Ray Conference}
\CopyRight{\copyright 2003 by Universal Academy Press, Inc.}
\pagenumbering{arabic}

\chapter{
Propagation of Light Elements in the Galaxy
}
\paperno

\author{%
%
%
Igor V.~Moskalenko,$^{1,2}$ Andrew W.~Strong,$^3$ Stepan G.~Mashnik,$^4$ Frank C.~Jones$^1$\\
{\it 
(1) NASA/Goddard Space Flight Center, Code 661, Greenbelt, MD 20771, USA\\
(2) JCA/University of Maryland, Baltimore County, Baltimore, MD 21250, USA\\
(3) MPI f\"ur extraterrestrische Physik, Postfach 1312, 85741 Garching, Germany\\
(4) Los Alamos National Laboratory, Los Alamos, NM 87545, USA
} \\
}

\section*{Abstract}
The origin and evolution of isotopes of the lightest elements H$^2$, He$^3$,
Li, Be, B in the universe is a key problem in such fields as
astrophysics of CR, Galactic evolution, non-thermal nucleosynthesis,
and cosmological studies.  One of the major sources of these species
is spallation by CR nuclei in the interstellar medium. On the other
hand, it is the B/C ratio in CR and Be$^{10}$ abundance which are
used to fix the propagation parameters and thus the spallation rate.  We
study the production and Galactic propagation of isotopes of elements $Z\leq5$ using the
numerical propagation code GALPROP and updated production cross
sections.

\section{Introduction}
It has been shown recently [9] that accurate measurements of the antiproton flux
made during the last solar minimum [12] pose a challenge to conventional
CR propagation models. A solution has been proposed [10] that the observed
CR may contain a fresh, local, ``unprocessed'' component at low energies (LE), perhaps
associated with the Local Bubble (LB). This component reduces the production
of B at LE allowing to fit the B/C ratio and antiproton flux simultaneously.
For more discussion see [11].

Isotopes of the lightest elements H$^2$, He$^3$,
Li, Be, B in CR are almost all secondary, being produced in spallations
of heavier nuclei on interstellar gas Galaxy-wide. Their abundances
and isotopic composition, therefore, might help to distinguish
between the propagation models. In this study we calculate 
abundances and isotopic composition of H, He, Li, Be, B in two
models, a conventional reacceleration model [9], and a model 
with a local component [10].

\section{The Model}
In our calculations we use the propagation model GALPROP (2D option)
as described elsewhere [9,14].
The nucleon injection spectrum of the Galactic CR is taken as
a modified power law in rigidity [6], for the injected 
particle density. The LB spectrum is taken as
a power law in rigidity with exponential cut off.
The Galactic halo size is fixed at 4 kpc.
To reduce possible errors due to the cross sections
we use our own fits to the data on reactions 
$p+\mathrm{He,C,N,O} \to \mathrm{H^2},\mathrm{He}^{3,4}$
and $p+\mathrm{C,N,O} \to \mathrm{Li,Be,B}$ that produce most of these
elements [8].
The heliospheric modulation is treated using the force-field
approximation; here we use $\Phi=500$ MV for all plots.

In the case of a conventional reacceleration model 
(model $\mathcal{A}$) we use
our standard methodologies: the propagation parameters
were derived from the fit to the B/C ratio, while source
abundances were tuned to ACE data [15] and HEAO-3 data [4]
at high energies (HE). No special tuning was done for antiprotons.

In the case of the model with a local component (model $\mathcal{B}$)
we use the following procedure. 
The HE part of the B/C ratio \emph{plus} antiproton 
flux measurements are used to restrict the value of the diffusion
coefficient and its energy dependence, while the LE part 
of the B/C ratio is used to fix the reacceleration level and 
define the parameters of the LB component. In this way
the model provides the best fit to all data at the cost of 
extra free parameters.

Model parameters are given in Table 1;
model $\mathcal{A}$ is similar to models DR/DRB as defined in [9], and
model $\mathcal{B}$ is similar to model DR II as defined in [10].
To better match the He spectrum (Fig.~1),
a major contributor to H$^2$ and He$^3$ production, in both models we 
introduced a break (at 14 GV in $\mathcal{A}$, and at 10 GV in $\mathcal{B}$) 
in He$^4$ (only) power-law injection spectrum.

\begin{table}[tb]
\caption{Propagation Parameter Sets}
\begin{center}
\begin{tabular}{cccccc}
\hline
\ 
   & Injection 
      & \multicolumn{2}{c}{Diffusion coeff.\ at 3GV} 
            & Alfv\'en speed
               & Source
\\
\cline{3-4}
Model
   & index, $\gamma$
      & $D_0$, cm$^2$ s$^{-1}$
         & Index, $\delta$
            &$v_A$, km s$^{-1}$
               & abundances
\\ 
\hline
  $\mathcal{A}$
& 1.94/2.42
& $6.25\times10^{28}$
& 0.33
& 36
& LE=HE     
\\
  $\mathcal{B}$
& 1.69/2.28
& $3.30\times10^{28}$
& 0.47
& 23
& LB$\neq$Galactic CR 
\\
\hline
\end{tabular}
\end{center}
\end{table}

\section{Results}
Fig.~2 shows the H$^2$/He$^4$ ratio. The measured ratio [2] is
a factor of $\sim$1.8 larger than expected. While
there are important hints that this may be caused 
by systematic effects, the analysis is still in progress
(D.\ Vasilas private comm.).

Fig.~3 shows other ratios.
Both models are consistent with data on isotopic ratios
of Li, Be, and B given the large error bars.
He$^3$/He$^4$ ratio is more sensitive.
Both models agree well with the data, however
to match the He$^3$/He$^4$ ratio \emph{and} He spectrum, model $\mathcal{B}$ requires 
a factor of $\sim$2 larger LB abundance of He$^4$.
The HE CR source abundance in both models He/Si $\approx100$,
while the LB abundance in model $\mathcal{B}$ is 220 (cf.\ solar system
value 2400). Taking into account that He and Si are abundant elements
a factor of 2 difference is significant. 
Contributions of C$^{12}$ and O$^{16}$
to He$^3$ and He$^4$ production appear to
be non-negligible.

The smaller proportion of He compared to heavier elements 
in Galactic CR sources supports the idea that
HE CR are accelerated in shocks from fresh SN ejecta.
The higher proportion of He compared to heavier elements
in model $\mathcal{B}$ at LE implies that the material was 
diluted before acceleration. 
This is in line with the view described in
[10] that the LB component may be accelerated by an
ensemble of weak shock waves out of the interstellar medium. 
If the model $\mathcal{B}$
is correct it may indicate that the LE part of the He and proton
Galactic spectra is flatter than thought with corresponding 
consequences for the diffuse Galactic $\gamma$-ray emission and
Galactic chemical evolution.

\begin{figure}[tb]
  \begin{center}
    \includegraphics[width=1\textwidth]{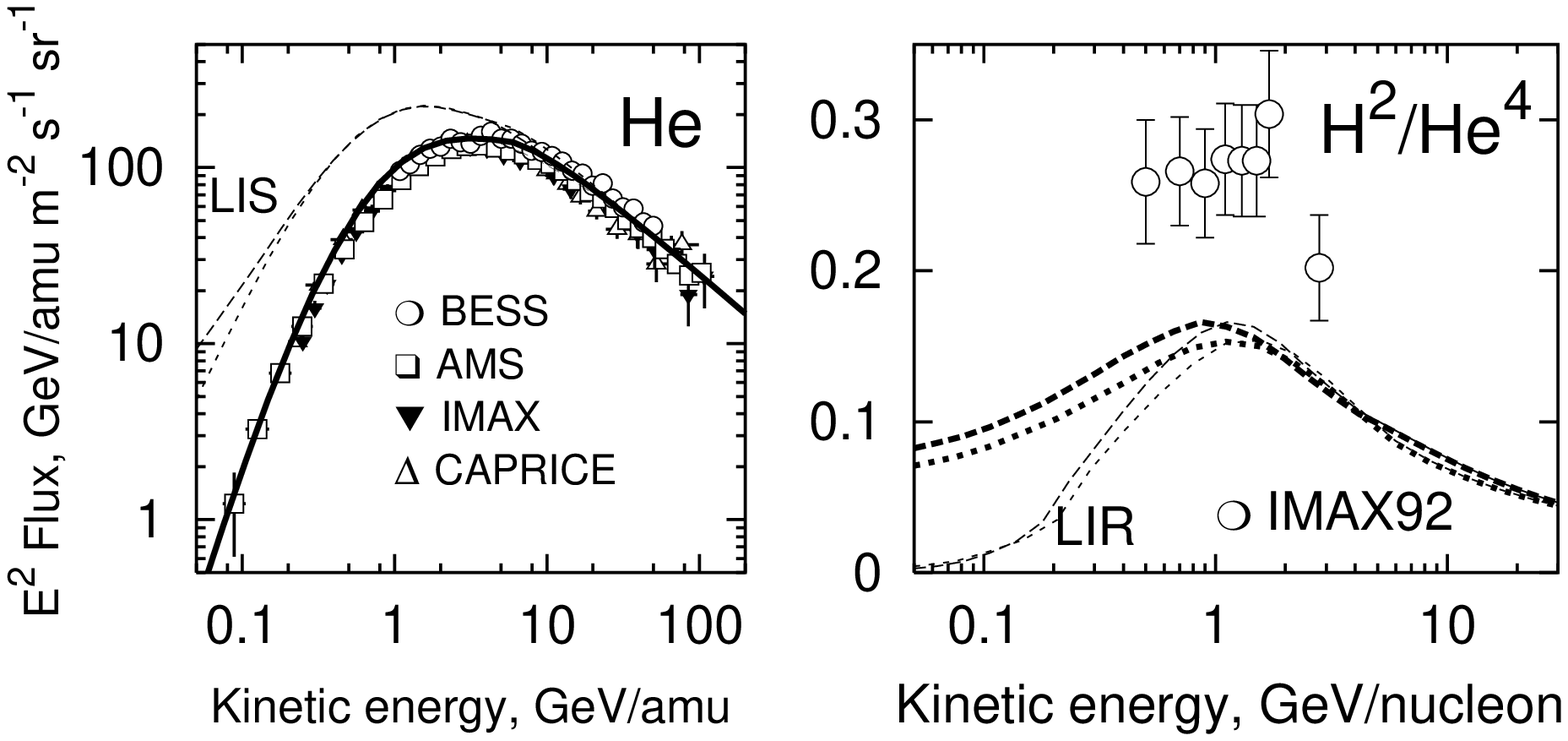}
  \end{center}
  \vspace{-0.5pc}
\begin{minipage}[tl]{0.48\textwidth}
  \caption{He spectrum. Calculations:
 \lower-2pt\hbox{\Large \bf ...} -- model $\mathcal{A}$, 
{\Large -\,-\,-} -- $\mathcal{B}$. Solid line -- modulated 
($\mathcal{A}$ \& $\mathcal{B}$), 
thin lines -- interstellar (LIS).
Data: for references see [9].
}
\end{minipage} 
\hfill
\begin{minipage}[tl]{0.48\textwidth}
  \caption{H$^2$/He$^4$ ratio. Calculations:
 \lower-2pt\hbox{\Large \bf ...} -- model $\mathcal{A}$, 
{\Large -\,-\,-} -- model $\mathcal{B}$. Bold lines -- modulated, 
thin lines -- local interstellar ratio (LIR).
Data: IMAX92 [2].
}
\end{minipage} 
\end{figure}

\begin{figure}[t]
  \begin{center}
    \includegraphics[width=1\textwidth]{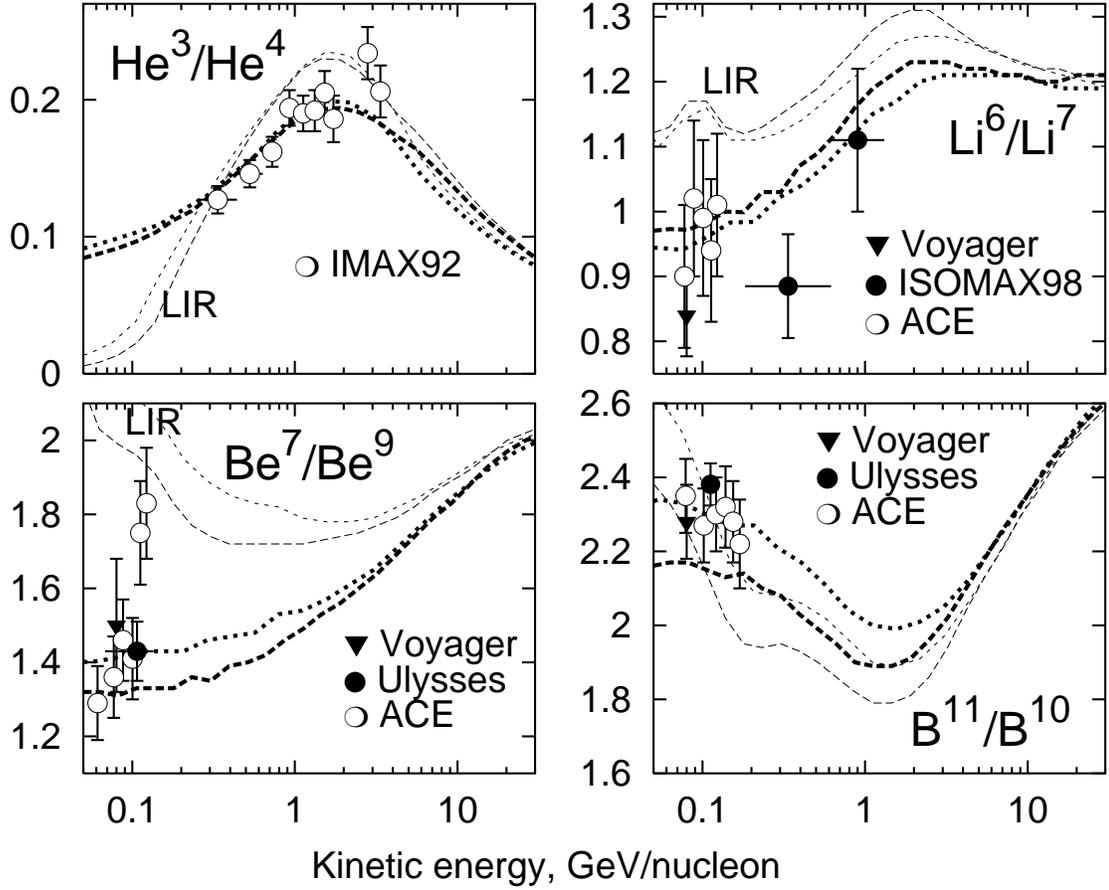}
  \end{center}
  \vspace{-0.5pc}
  \caption{Isotopic ratios as calculated in two propagation models.
Lines are coded as in Fig.\ 2. 
Data: IMAX92 [13], ISOMAX98 [5] (with statistical errors only), Voyager [7], 
Ulysses [1], ACE [3].
}
\end{figure}

This work was supported in part by 
a NASA Astrophysics Theory Program grant
and by the US Department of Energy.

\section{References}
\re
1.\ Connell J.J.\ 1998, ApJ 501, L59
\re
2.\ de Nolfo G.A.\ et al.\ 2000, in Proc.\ ACE-2000 Symp., ed.\ 
    Mewaldt R.A.\ et al.\ (AIP: NY), AIP 528, 425 
\re
3.\ de Nolfo G.A.\ et al.\ 2001, in Proc.\ 27th ICRC (Hamburg), 1667 
\re
4.\ Engelmann J.J.\ et al.\ 1990, A\&A 233, 96 
\re
5.\ G\"obel H.\ et al.\ 2001, in Proc.\ 27th ICRC (Hamburg), 1663 
\re 
6.\ Jones F.C., Lukasiak A., Ptuskin V., Webber W.\ 2001, ApJ 546, 264
\re  
7.\ Lukasiak A., McDonald F.B., Webber W.R.\ 1999, in Proc.\ 26th ICRC (Salt Lake City), 3, 41
\re
8.\ Moskalenko I.V., Mashnik S.G.\ 2003, these Proc.
\re
9.\ Moskalenko I.V., Strong A.W., Ormes J.F., Potgieter M.S.\ 2002, ApJ 565, 280
\re
10.\ Moskalenko I.V., Strong A.W., Mashnik S.G., Ormes J.F.\ 2003, ApJ 586, 1050
\re
11.\ Moskalenko I.V., Strong A.W., Mashnik S.G., Ormes J.F.\ 2003, these Proc.
\re
12.\ Orito S.\ et al.\ 2000, Phys.\ Rev.\ Lett.\ 84, 1078 
\re
13.\ Reimer O.\ et al.\ 1998, ApJ 496, 490 
\re
14.\ Strong A.W., Moskalenko I.V.\ 1998, ApJ 509, 212 
\re
15.\ Wiedenbeck M.E.\ et al.\ 2001, Spa.\ Sci.\ Rev.\ 99, 15 

\endofpaper
\end{document}